\documentclass[paper, twocolumn]{aastex631}
\usepackage{graphicx} 
\usepackage{threeparttable}
\usepackage{multirow}
\usepackage{color}
\usepackage{amsmath}
\usepackage{breakcites}

\newcommand\red{\textcolor{black}}

\submitjournal{ApJ}

\begin{document}
\title{Global deceleration and inward movements of X-ray knots and rims of RCW~103}

\shorttitle{Deceleration of X-ray knots and rims of RCW 103}
\shortauthors{Suzuki et al.}

\correspondingauthor{H. Suzuki}
\email{hiromasa050701@gmail.com}

\author[0000-0002-8152-6172]{Hiromasa Suzuki}
\affiliation{Department of Physics, Faculty of Science and Engineering, Konan University, 8-9-1 Okamoto, Higashinada, Kobe, Hyogo 658-8501, Japan}

\author[0000-0002-4383-0368]{Takaaki Tanaka}
\affiliation{Department of Physics, Faculty of Science and Engineering, Konan University, 8-9-1 Okamoto, Higashinada, Kobe, Hyogo 658-8501, Japan}

\author[0000-0002-7935-8771]{Tsuyoshi Inoue}
\affiliation{Department of Physics, Faculty of Science and Engineering, Konan University, 8-9-1 Okamoto, Higashinada, Kobe, Hyogo 658-8501, Japan}

\author[0000-0003-1518-2188]{Hiroyuki Uchida}
\affiliation{Department of Physics, Kyoto University, Kitashirakawa Oiwake-cho, Sakyo, Kyoto 606-8502, Japan}

\author[0009-0006-7889-6144]{Takuto Narita}
\affiliation{Department of Physics, Kyoto University, Kitashirakawa Oiwake-cho, Sakyo, Kyoto 606-8502, Japan}


\begin{abstract}

Kinematics of shocks, ejecta knots, and the compact remnant of a supernova remnant gives an insight into the nature of the progenitor and surrounding environment.
We report on a proper motion measurement of X-ray knots and rims of the magnetar-hosting supernova remnant RCW~103.
Chandra data obtained in three epochs, 1999, 2010, and 2016 are used.
We find a global deceleration of 12 knots and rims both in northern and southern regions within the last $\sim 24$~yrs, even though its age is thought to be larger than 2~kyr.
Some of them even changed their moving directions from outward ($\sim 1,000$~km~s$^{-1}$) to inward ($\sim -2,000$~km~s$^{-1}$).
Our findings can be explained with a collision with a high-density medium both in the northern and southern edges of the remnant, although the remnant may still be expanding in the wind-blown cavity.
The proper motion of the associated magnetar 1E~161348$-$5055 is possibly detected with a velocity of $\approx 500$~km~s$^{-1}$.

\end{abstract}

\section{Introduction}
RCW~103 is a young or middle-aged supernova remnant (SNR) hosting the compact object 1E~161348$-$5055 \citep{tuohy80}.
Its age, i.e., elapsed time after the supernova explosion, is estimated to be 2.0--4.4~kyr \citep{carter97, braun19}.
It has a nearly circular shape with a spatial extent of $\sim 10\arcmin$ or $\sim 9$~pc at the estimated distance of 3.1~kpc \citep{reynoso04}.
Interestingly, its morphologies are similar among radio, infrared, optical, and X-rays.
All show bright emissions in the southern large area and northern small part.
Radio continuum observations revealed a smooth structure without clear shells \citep{dickel96}.
\cite{paron06} found an interacting $^{12}$CO cloud in the southern area.
Infrared observations also found interacting H$_2$ gas and other elements \citep{oliva90, oliva99, rho01, reach06, pinheiro11}.
A 1720~MHz OH maser detection from the southern area also supports the cloud interaction \citep{frail96}.
\cite{carter97} detected H$\alpha$ filaments from both south and north, with the northern filament being much fainter.
They estimated the age to be $\sim 2$~kyr based on optical proper motions of $\sim 1,100$~km~s$^{-1}$.

The compact object 1E~161348$-$5055 has been known as an extraordinary compact object with a very long periodicity $\sim 6.67$~h \citep{deluca06}.
In 2016, it exhibited a bursting activity and began to be recognized as a magnetar \citep{dai16, rea16, tendulkar17}.
Previous X-ray observations shed light on the relation between the progenitor and magnetar \citep{nugent84, frank15, braun19, zhou19}.
A common conclusion is that the supernova explosion was less energetic (with an explosion energy of $10^{49}$--$10^{50}$~erg) and the progenitor was not very massive ($\lesssim 13$~M$_{\odot}$).
Most recently, \cite{narita23} identified X-ray emission from shock-heated circumstellar medium (CSM) near the edges of RCW~103. They found an enhanced N/O abundance ratio ($\sim 4$) of the CSM, and suggested that the progenitor rotation was not rapid ($\lesssim 100$~km~s$^{-1}$) and a magnetar formation by dynamo effects in massive stars \red{($> 35$~M$_{\odot}$)} is unlikely.

From another aspect, constraining the X-ray kinematic properties including movements of forward shocks, ejecta knots, and the associated magnetar is of great importance as well to understand the nature of the progenitor and magnetar.
In this paper, we report on proper motion measurements of X-ray bright knots and rims, and the associated magnetar.
Out original purpose was to determine the explosion center and obtain tight constraints on the age and kinematics.
However, we find a global deceleration and inward movements of the X-ray knots and rims.
In Section~\ref{sec-obs}, we summarize the observation log and data reduction processes.
Our proper motion analysis and results are described in Section~\ref{sec-ana}.
We discuss the origin of the deceleration and inward movements in Section~\ref{sec-dis}, and conclude in Section~\ref{sec-con}.

\section{Observation and data reduction}\label{sec-obs}
We use five Chandra ACIS-I \citep{garmire97} observations of the RCW~103 region listed in Table~\ref{tab-obs}, which consist of three epochs (1999, 2010, and 2016).
The baselines for the proper motion study are $\approx 11$ yr and $\approx 6$ yr, for the first and second intervals, respectively.
The observation log is summarized in Table~\ref{tab-obs}.

We use the analysis software CIAO (v4.15; \citealt{fruscione06}) and calibration database v4.10.2 for the data reduction and analysis.
We process the raw data using the standard data reduction method ({\tt chandra\_repro}).

\begin{table*}[htb!]
\centering
\caption{Chandra observation log
\label{tab-obs}}
\begin{tabular}{l l l l l l l l}
\hline\hline
ObsID & R.A. (2000) & Dec. (2000) & Date & Exposure (ks) & $\Delta x$ (pixel)\tablenotemark{a} & $\Delta y$ (pixel)\tablenotemark{a} \\ \hline
123 & 244$\fdg$40550 & 51$\fdg$02022 & 1999 Sep. 26 & 13.4 & 0.51 & 0.39 \\
11823 & 244$\fdg$41094 & 51$\fdg$02284 & 2010 Jun. 01 & 62.5 & \nodata & \nodata \\
12224 & 244$\fdg$40770 & 51$\fdg$02419 & 2010 Jun. 27 & 17.8 & $-0.10$ & 0.13 \\
18459 & 244$\fdg$41579 & 51$\fdg$04894 & 2016 May 23 & 25.8 & \nodata\tablenotemark{b} & \nodata\tablenotemark{b} \\
18854 & 244$\fdg$41333 & 51$\fdg$04969 & 2016 May 30 & 13.0 & $-0.64$ & 0.29 \\ \hline
\end{tabular}
\tablenotetext{a}{Coordinate transformation parameters with respect to those of ObsID 11823. Transformation directions $+\Delta x$ and $+\Delta y$ correspond to $-$R.A. and $+$Dec., respectively.}
\tablenotetext{b}{No correction is performed due to the low quality of point-like sources.}

\end{table*}

\section{Analysis and results}\label{sec-ana}
We perform a proper motion study on RCW~103. The procedures and results are presented in this section.
In our analysis, we use the software HEASoft (v6.30.1; \citealt{heasarc14}), XSPEC (v12.12.1; \citealt{arnaud96}), and AtomDB 3.0.9.
Throughout the paper, uncertainties in the text, figures, and tables indicate $1\sigma$ confidence intervals.

\begin{figure}[htb!]
\centering
\includegraphics[width=8cm]{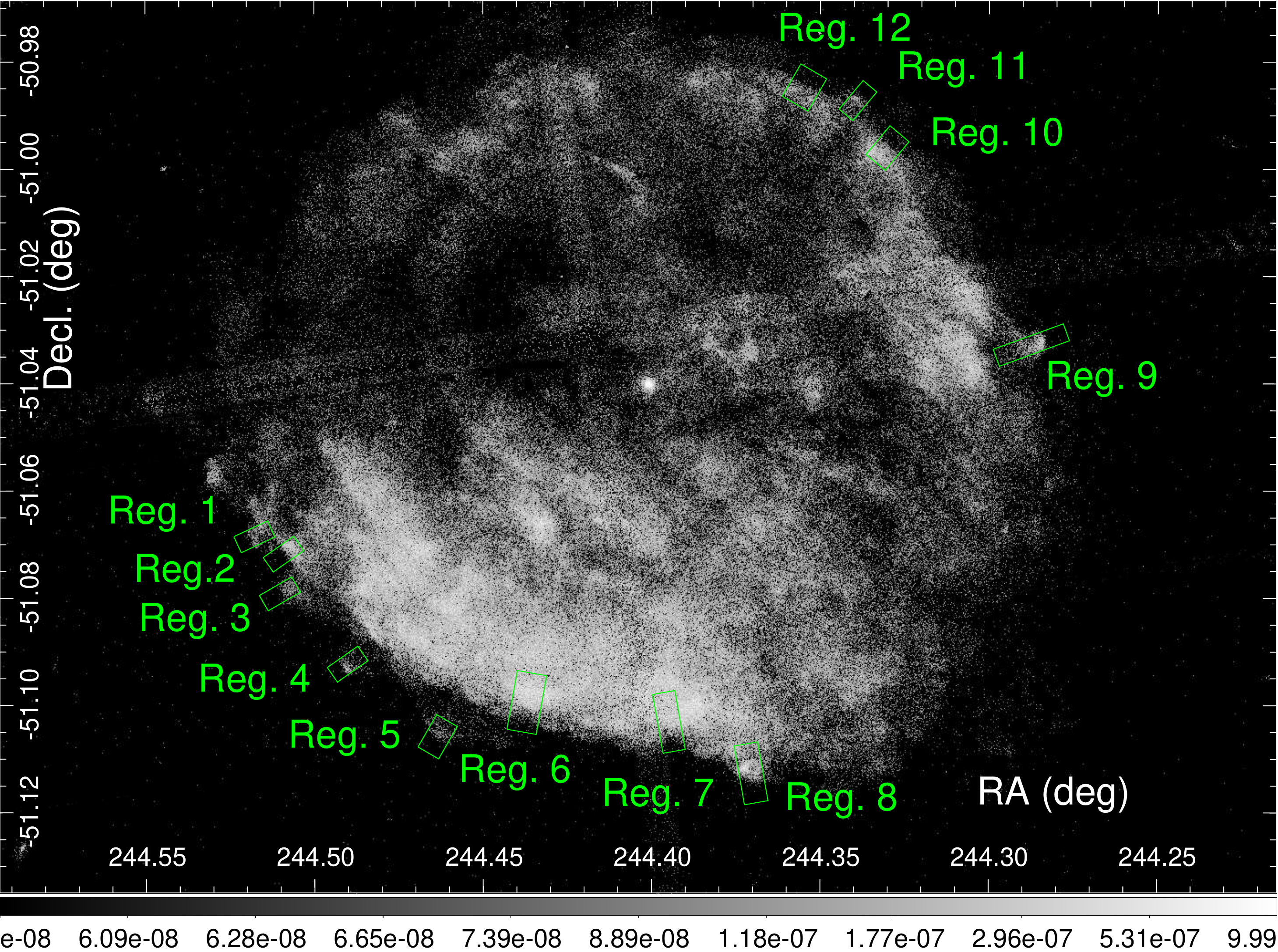}
\caption{Exposure-corrected Chandra image of RCW~103 in the energy band of 0.5--5.0 keV. The radial-profile extraction regions are indicated with the green boxes.
\label{fig-image}}
\end{figure}

Figure~\ref{fig-image} presents an X-ray image of the whole remnant with our analysis regions.
We choose bright knots and sharp edges as our analysis regions.
The profile extraction directions are determined by eye to roughly correspond to the directions perpendicular to the boundaries or toward the geometric center of the remnant.

\subsection{Aspect correction}
To maximize the reliability and accuracy of the proper motion measurement, we perform the aspect correction to individual observations.
As a large fraction of the central region of the field of view is covered by the bright target source, we only find 7--9 point-like sources with off-axis angles of $>4\arcmin$. We perform the correction with the CIAO tool {\tt wcs\_match} and {\tt wcs\_update}. Considering the small number of detected point-like sources and their off-axis positions, we perform coordinate transformations without rotation and scaling.
The resultant transformation parameters are listed in Table~\ref{tab-obs}.
The relative offsets of 0.1--0.6~pixels, which correspond to 0$\farcs$05--0$\farcs$3, are reasonable according to the pointing accuracy of Chandra ACIS-I, $\approx 0\farcs$67.\footnote{The reference for the pointing accuracy is \url{https://cxc.harvard.edu/cal/ASPECT/celmon/}.}

\subsection{Proper motions of X-ray knots and rims}
\begin{figure*}[htb!]
\centering
\includegraphics[width=14cm]{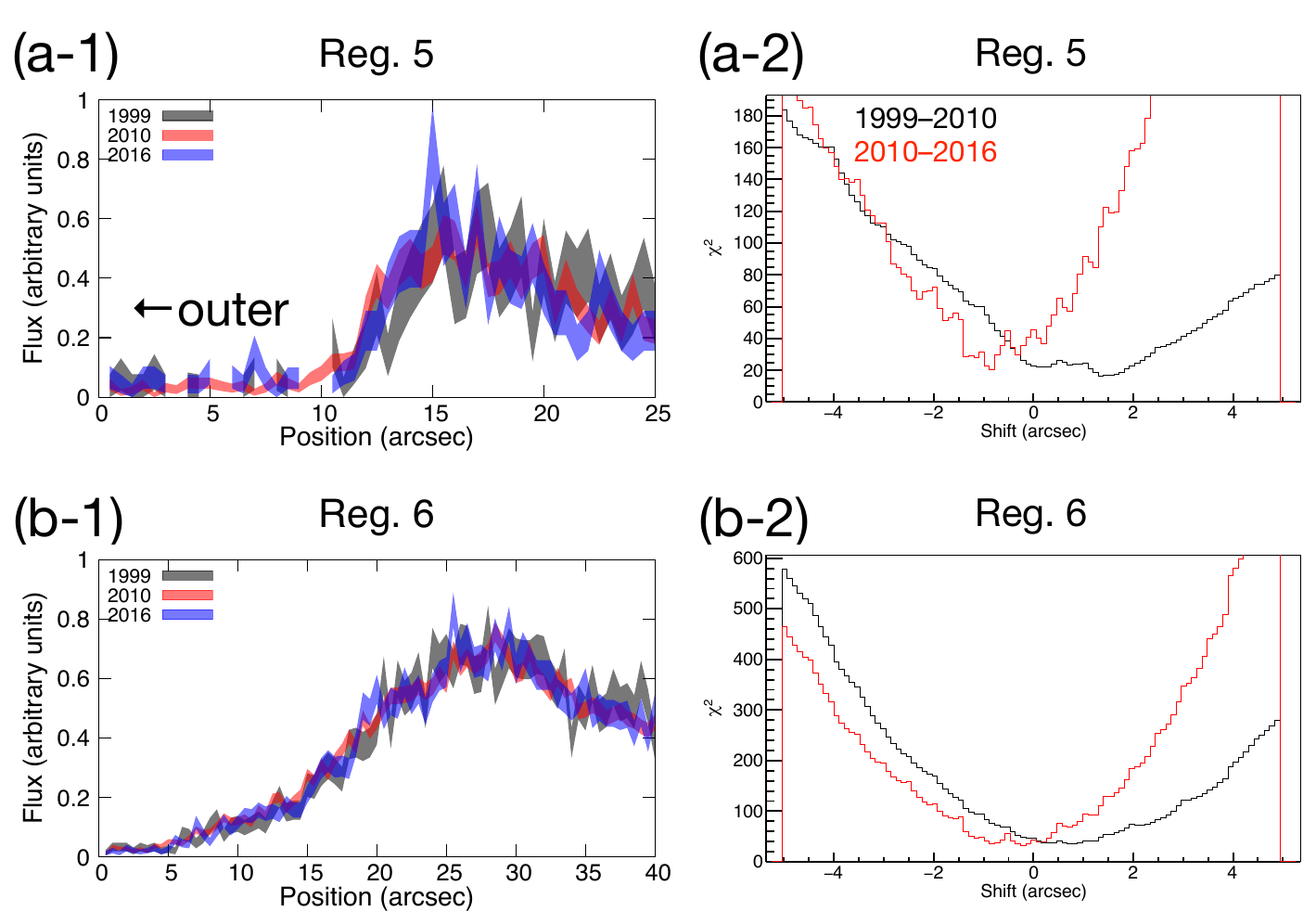}
\caption{Radial profiles and derived $\chi^2$--$\Delta x$ (shift) plots extracted from Reg. 5 (a) and from Reg. 6 (b).
In the radial profiles, negative directions correspond to outer regions. In the $\chi^2$--$\Delta x$ plots, positive values indicate outward movements.
\label{fig-profile}}
\end{figure*}

We measure proper motions using radial profiles extracted from the regions indicated in Figure~\ref{fig-image}.
The two observations in 2010 are merged after the aspect correction. The same process is done for the observations in 2016.
Thus, hereafter, we use three images obtained in 1999, 2010, and 2016.
Flux profiles are extracted from the exposure-corrected images \red{in the 0.5--5.0~keV energy range}.
Two examples are shown in Figure~\ref{fig-profile} (a-1) and (b-1).

We use the same method as that taken in \cite{tanaka21} and \cite{suzuki22b} in calculating the velocities.
Two profiles obtained from two epochs are used.
We artificially shift the second profile by $\Delta x$ and evaluate the difference against the first one with $\chi^2 (\Delta x)$, which is defined as
\begin{linenomath}
\begin{equation}
\chi^2 (\Delta x) = \sum_{i} \frac{(f_{i} - g_i (\Delta x))^2}{\Delta f_{i}^2 + \Delta g_i (\Delta x)^2},
\end{equation}
\end{linenomath}
where $f_i$ and $\Delta f_i$ indicate the flux and error of the bin number $i$ in the first observation, and $g_i$ and $\Delta g_i$ indicate those of the shifted second profile.
This calculation is repeated with various values of $\Delta x$ and we obtain $\chi^2$ as a function of $\Delta x$.
The minimum $\chi^2$ value ($\chi^2_{\rm min}$) and corresponding shift ($\Delta x_{\rm min}$) are determined by fitting the $\chi^2$--$\Delta x$ plot with a parabola function.
We calculate proper motion velocity from the best-fit $\Delta x_{\rm min}$ and known baselines.
The profile shift is not limited to an integer multiple of the bin width.
We re-bin the shifted profile $g (\Delta x)$ with the same bin arrangement as $f$ with an assumption of a uniform probability distribution inside each bin.
Then, the profile-shift ranges that give $\chi^2 (\Delta x) = \chi^2_{\rm min} +1$ are calculated from the best-fit parabola functions.
These ranges are considered to be 1$\sigma$ confidence ranges of the profiles shifts.

\begin{figure}[htb!]
\centering
\includegraphics[width=8cm]{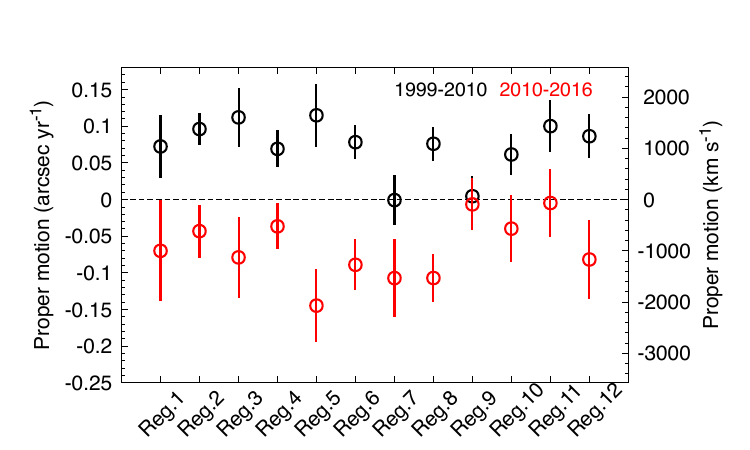}
\caption{Proper motion velocities of our analysis regions. Positive velocities indicate outward movements.
A distance of 3.1~kpc is assumed in calculation of the velocities.
\label{fig-velocity}}
\end{figure}

The $\chi^2$--$\Delta x$ plots of Reg.~5 and Reg.~6 are shown in Figure~\ref{fig-profile} (a-2) and (b-2), respectively.
The $\Delta x_{\rm min}$ values in the second interval (2010 to 2016) are negative, whereas those in the first (1999 to 2010) are positive.
These indicate that their movements were outward before 2010 but they changed their moving directions to inward after 2010.
We show the calculated velocities of all the analysis regions in the two intervals in \red{Table~\ref{tab-prop} and} Figure~\ref{fig-velocity}.
One can see a global deceleration from the first to second interval. Among them, Regs.~5--8 are \red{firmly} found to be moving inward in the second interval.

We here evaluate possible systematic uncertainties.
The pointing accuracy of Chandra has to be considered, which would be $< 0\farcs67$ after the aspect correction.
Even if we assume that the astrometry offsets between the images are significant, some of the knots or rims still should be moving inward in the second interval, because the analysis regions include both northern and southern edges.
The profile extraction directions are determined by eye, and thus the measured velocities will have some uncertainties due to deviations from the true moving directions.
We evaluate this uncertainty by slightly changing the extraction direction ($\pm10$~deg) of Reg.~6, finding a $\lesssim 100$~km~s$^{-1}$ variation in velocities. This uncertainty is not significant because the statistical uncertainties are much larger.
\red{We also repeat the analysis procedure with 1) an alternative aspect correction with an optical source catalog and 2) different extraction energy ranges of 1.0--5.0~keV and 1.5--5.0~keV. The measured velocities are largely consistent with the ones obtained above with the same tendency (See Appendix~\ref{sec-catalog} and \ref{sec-highE}).}

\begin{table}[htb!]
\centering
\caption{\red{Proper motion velocities\tablenotemark{a}}
\label{tab-prop}}
\begin{tabular}{l r r }
\hline\hline
Region & Velocity (km~s$^{-1}$) & Velocity (km~s$^{-1}$) \\ 
& 1999--2010 & 2010--2016 \\
\hline
Reg.1 & $1000 \pm 600$ & $-1000 \pm 1000$ \\
Reg.2 & $1400 \pm 300$ & $-600 \pm 500$ \\
Reg.3 & $1600 \pm 600$ & $-1100 \pm 800$ \\
Reg.4 & $990 \pm 400$ & $-500 \pm 400$ \\
Reg.5 & $1600 \pm 600$ & $-2100 \pm 700$ \\
Reg.6 & $1100 \pm 300$ & $-1300 \pm 500$ \\
Reg.7 & $-10 \pm 500$ & $-1500 \pm 800$ \\
Reg.8 & $1100 \pm 300$ & $-1500 \pm 500$ \\
Reg.9 & $60 \pm 400$ & $-100 \pm 500$ \\
Reg.10 & $880 \pm 400$ & $-600 \pm 600$ \\
Reg.11 & $1400 \pm 500$ & $-70 \pm 700$ \\
Reg.12 & $1200 \pm 400$ & $-1200 \pm 800$ \\
\hline
\end{tabular}
\tablenotetext{a}{A distance of 3.1~kpc is assumed. Minus velocities indicate inward movements (Same for Table~\ref{tab-prop-catalog} and \ref{tab-prop-highE}).}
\end{table}

\subsection{Proper motion of the associated magnetar 1E~161348$-$5055}\label{sec-cco}
\begin{figure}[htb!]
\centering
\includegraphics[width=8cm]{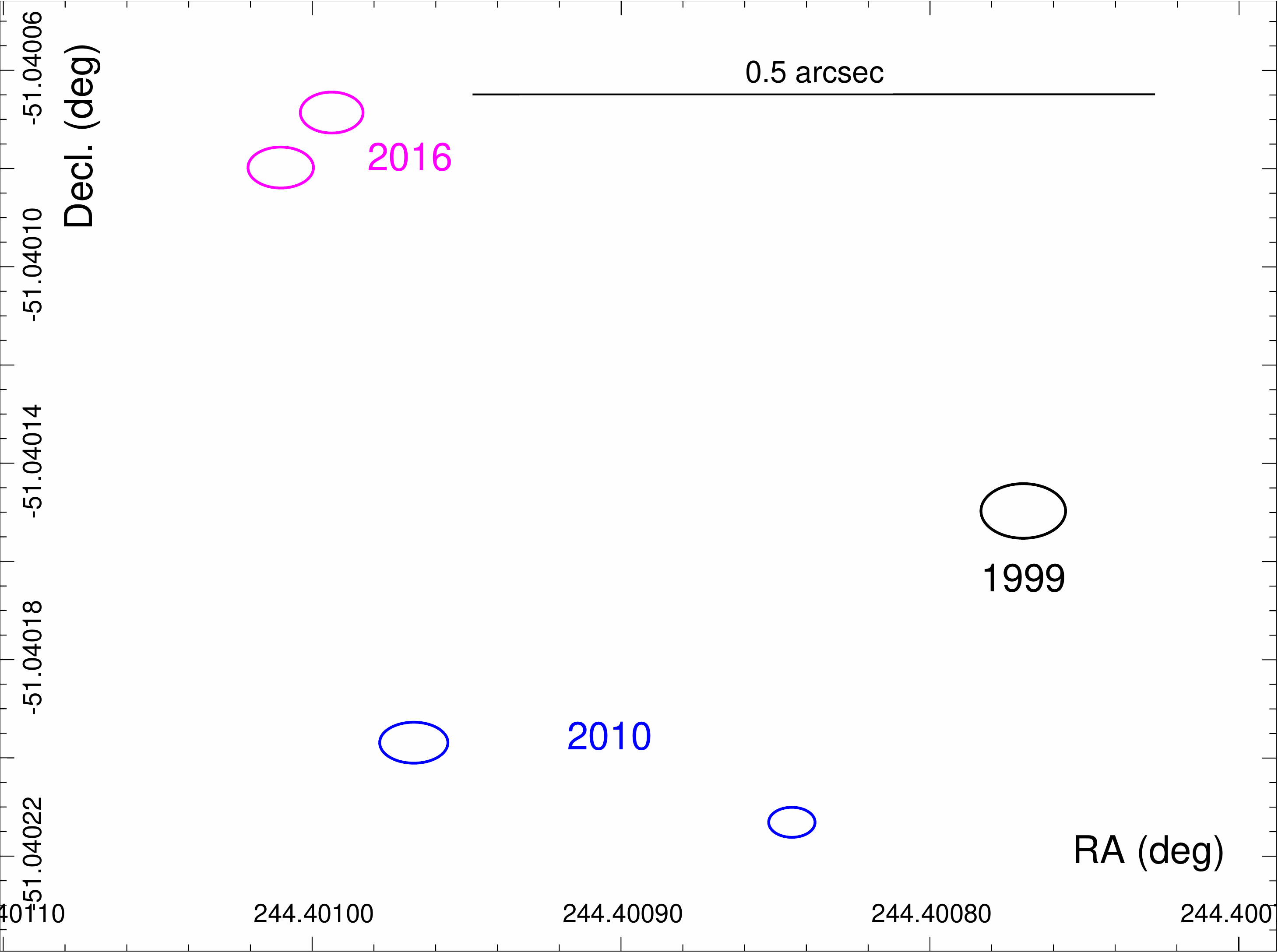}
\caption{Locations of the magnetar 1E~161348$-$5055 in 1999, 2010, and 2016. The central positions and radii of the ellipses indicate the estimated positions of the magnetar at different times and their statistical errors.
\label{fig-cco}}
\end{figure}

Using the aspect-corrected, \red{0.5--5.0~keV} images, we measure the proper motion of the associated magnetar, 1E~161348$-$5055.
For individual observations, we determine the positions of the magnetar and their statistical errors using the CIAO tool {\tt wavdetect}.
The determined locations are presented in Figure~\ref{fig-cco}.
The angular displacement between 1999 and 2016 (ObsID 18854) is measured to be $0\farcs585 \pm 0\farcs025$, which is converted to $501 \pm 21$~km s$^{-1}$ at a distance of 3.1~kpc \citep{reynoso04}.
We note that this displacement might be insignificant if we consider systematic uncertainties due to the pointing accuracy.

\section{Discussion}\label{sec-dis}
We find a global deceleration of the X-ray knots and rims in RCW~103 in the last $\sim 24$~yrs, even though its age is thought to be larger than 2~kyr.
Some of them were even moving inward in the second interval, from 2010 to 2016.
We here discuss the origin of this sudden deceleration.

\cite{narita23} proposed that X-ray emitting plasma near the outer edges are CSM dominated. They also suggested that the remnant is still expanding in the wind-blown bubble based on the derived progenitor properties.
The X-ray bright southern and northern edges, on which this work focuses, coincide with the locations of H$\alpha$ emission \citep{carter97}.
The southern edge is thought to be interacting with a molecular cloud \citep{dickel96}. Considering these facts and suggestions, we propose a scenario that both northern and southern regions interact with molecular or atomic clouds, although the remnant is still expanding in the wind-blown bubble. We assume that the northern part is also interacting with a high-density medium but it is yet to be detected. The weaker H$\alpha$ emission and slower deceleration in the northern part support an interpretation that the interacting medium there has a lower density than the southern part.

Regs.~5--8 are found to have decelerated from $\sim +1,000$~km~s$^{-1}$ (outward) to $\sim -2,000$~km~s$^{-1}$ (inward).
This can be interpreted as a reflection of the shocks due to a collision with a high-density medium.
The shock reflection by an interaction with a high density cloud is studied analytically by \cite{MZ94} and \cite{IY12}.
For a high Mach number incident shock, like an SNR blast wave shock, the relation between the incident/reflection shock velocities and density jump at a cloud surface is given by eq.~(A3) of \cite{IY12}.
In the case of an incident shock velocity $\sim 1,000$~km s$^{-1}$ and a reflection shock velocity $\sim -2,000$~km s$^{-1}$, the required density jump is calculated to be $\sim 36$.
This is consistent with a typical density jump between a diffuse ISM and HI clouds.
\red{We note that a shock wave can be reflected whereas shock-heated plasma will not be. The reflected shock enhances thermal X-rays while moving inward, which can be observed as an inward movement if the newly enhanced emission is bright enough. If the observed X-ray radial profiles originate from mixtures of outward- and inward-moving plasma, actual reflected-shock velocities in the observer's frame might be larger than the measured velocities.}

Inward moving filaments were found in a few other young SNRs, such as Cassiopeia~A \citep{sato18} and RCW~86 \citep{suzuki22b}. The inward filaments are interpreted as reverse shocks for Cassiopeia~A and reflection shocks for RCW~86. The case of RCW~103 is similar to RCW~86. A large difference is the locations where the inward movements are observed: in the present case, they are at the outer edges of the X-ray emission, whereas they are well behind outermost filaments in the case of RCW~86. This is consistent with our interpretation, a very recent collision in RCW~103.

To summarize, the global deceleration can be understood as a result of a collision of the shocks with a high-density medium (molecular or atomic cloud), although the X-ray emitting plasma may still be expanding in the wind-blown bubble.

\section{Conclusion}\label{sec-con}
We examined proper motions of X-ray knots and rims in the southern and northern edges of RCW~103.
We found a global deceleration of them within the last $\sim 24$~yrs, even though its age is thought to be larger than 2~kyr.
Among them, Regs.~5--8 were found to have changed the moving directions from $\sim +1,000$~km~s$^{-1}$ (outward) to $\sim -2,000$~km~s$^{-1}$ (inward).
We confirmed that the deceleration and inward movements are robust to the uncertainties in the moving directions and the pointing accuracy of Chandra.
The inward movements can be understood as a shock reflection due to a collision with a high-density medium.
As a conclusion, the global deceleration can be explained as due to a collision with a high-density medium both in the northern and southern regions, although the X-ray emitting plasma may still be expanding in the wind-blown bubble.

\begin{acknowledgments}
We appreciate a fruitful discussion with H. Sano about the surrounding medium.
This research has made use of the VizieR catalogue access tool, CDS, Strasbourg, France (DOI : 10.26093/cds/vizier). The original description of the VizieR service was published in 2000, A\&AS 143, 23.
This work was partially supported by JSPS/MEXT grant Nos.~JP21J00031 (HS), JP19H01936, JP21H04493 (TT), and JP22H01265 (HU).

\end{acknowledgments}

\vspace{5mm}
\facilities{Chandra
}

\software{HEASoft (v6.30.1; \citealt{heasarc14}), CIAO (v4.15; \citealt{fruscione06})
}


\appendix
\section{Aspect correction with the NOMAD-1 optical source catalog}\label{sec-catalog}
\red{In order to evaluate systematic uncertainties associated with the aspect correction, we here apply another aspect correction. We use the NOMAD-1 optical source catalog \citep{zacharias04} available via the VizieR service\footnote{\url{https://doi.org/10.26093/cds/vizier}} \citep{ochsenbein00} to register the point-like sources in the Chandra images. We find 4--6 X-ray sources (depending on observations) which match the catalog sources. After correcting all the images, we measure the proper motions in the same way as in Section~\ref{sec-ana}. The resultant velocities are listed in Table~\ref{tab-prop-catalog}. Overall, the velocities are consistent with the ones obtained in Section~\ref{sec-ana}, suggesting that the systematic uncertaintes due to the aspect correction are small compared to the statistical errors.}

\begin{table}[htb!]
\centering
\caption{\red{Proper motion velocities in the alternative aspect correction case}
\label{tab-prop-catalog}}
\begin{tabular}{l r r }
\hline\hline
Region & Velocity (km~s$^{-1}$) & Velocity (km~s$^{-1}$) \\ 
& 1999--2010 & 2010--2016 \\
\hline
Reg.1 & $1300 \pm 500$ & $-3000 \pm 1000$ \\
Reg.2 & $1100 \pm 300$ & $-1500 \pm 500$ \\
Reg.3 & $1100 \pm 600$ & $-2100 \pm 800$ \\
Reg.4 & $900 \pm 400$ & $-1600 \pm 500$ \\
Reg.5 & $1700 \pm 600$ & $-2400 \pm 700$ \\
Reg.6 & $1300 \pm 300$ & $-900 \pm 500$ \\
Reg.7 & $400 \pm 500$ & $-1300 \pm 700$ \\
Reg.8 & $1400 \pm 300$ & $-400 \pm 500$ \\
Reg.9 & $600 \pm 400$ & $1400 \pm 600$ \\
Reg.10 & $900 \pm 400$ & $-100 \pm 700$ \\
Reg.11 & $800 \pm 500$ & $400 \pm 700$ \\
Reg.12 & $1500 \pm 400$ & $-900 \pm 800$ \\
Reg.13 & $100 \pm 800$ & $-2900 \pm 1200$ \\
\hline
\end{tabular}
\end{table}

\section{Proper motions in different energy ranges}\label{sec-highE}
\red{We here check for systematic uncertainties of the proper motions due to the extraction energy range.
Because the detector calibration might be less reliable at low energies due to the contamination on the sensor surface \citep{marshall04, odell15, plucinsky18}, we test two additional cases where we use the 1.0--5.0~keV and 1.5--5.0~keV energy ranges. The measured velocities are listed in Table~\ref{tab-prop-highE}.
In the latter case, velocities are constrained only for Regs.~6, 7, and 8, due to the limited statistics.
In both cases, one can see that the results are mostly consistent with the ones in Section~\ref{sec-ana}, showing a global deceleration and a change in moving directions.}

\begin{table}[htb!]
\centering
\caption{\red{Proper motion velocities in the different energy-selection cases}
\label{tab-prop-highE}}
\begin{tabular}{l r r }
\hline\hline
Region & Velocity (km~s$^{-1}$) & Velocity (km~s$^{-1}$) \\ 
& 1999--2010 & 2010--2016 \\
\hline
\multicolumn{2}{l}{1.0--5.0~keV case} \\
Reg.1 & $600 \pm 1100$ & $-2000 \pm 1100$ \\
Reg.2 & $2200 \pm 500$ & $-1000 \pm 600$ \\
Reg.3 & $800 \pm 900$ & $-300 \pm 1000$ \\
Reg.4 & $-500 \pm 900$ & $-600 \pm 600$ \\
Reg.5 & $1400 \pm 900$ & $-1400 \pm 900$ \\
Reg.6 & $1700 \pm 400$ & $-1309 \pm 600$ \\
Reg.7 & $100 \pm 600$ & $-1600 \pm 900$ \\
Reg.8 & $1200 \pm 400$ & $-1600 \pm 500$ \\
Reg.9 & $400 \pm 500$ & $-400 \pm 600$ \\
Reg.10 & $800 \pm 500$ & $-500 \pm 800$ \\
Reg.11 & $1700 \pm 600$ & $100 \pm 800$ \\
Reg.12 & $1000 \pm 600$ & $-500 \pm 900$ \\
Reg.13 & $1400 \pm 1100$ & $-3200 \pm 1700$ \\
 \hline
\multicolumn{2}{l}{1.5--5.0~keV case} \\
Reg.6 & $2100 \pm 900$ & $-3000 \pm 1100$ \\
Reg.7 & $1700 \pm 2000$ & $-2300 \pm 1500$ \\
Reg.8 & $900 \pm 1000$ & $-1100 \pm 900$ \\
\hline
\end{tabular}
\end{table}

\bibliography{references.bib}
\bibliographystyle{aasjournal}


\end{document}